\documentclass[runningheads,a4paper]{llncs}

\usepackage{amssymb}
\usepackage{graphicx}
\usepackage{xcolor}
\usepackage{ccaption}
\usepackage[CaptionAfterwards]{fltpage}
\usepackage{tabularx}
\usepackage{rotating}

\usepackage{url}
\urldef{\mailsa}\path|stys@jcu.cz|
\newcommand{\keywords}[1]{\par\addvspace\baselineskip
\noindent\keywordname\enspace\ignorespaces#1}

\begin{document}

\mainmatter  % start of an individual contribution

\title{Optimal noise in the hodgepodge machine simulation of the Belousov-Zhabotinsky reaction}

\titlerunning{Optimal noise in hodgepodge machine}

\author{Dalibor \v{S}tys \and Kry\v{s}tof M. \v{S}tys \and Anna Zhyrova \and Renata Rycht\'{a}rikov\'{a}}
\authorrunning{\v{S}tys et al.}

\institute{University of South Bohemia in \v{C}esk\'{e} Bud\v{e}jovice, Faculty of Fisheries and Protection of Waters, South Bohemian Research Center of Aquaculture and Biodiversity of Hydrocenoses, Institute of Complex Systems, Z\'{a}mek 136, 373 33 Nov\'{e} Hrady, Czech Republic\\
\mailsa\\
\url{http://www.frov.jcu.cz/cs/ustav-komplexnich-systemu-uks}}


\toctitle{Optimal noise in hodgepodge machine}
\tocauthor{Stys et al.}
\maketitle

\begin{abstract}
One of the simplest multilevel cellular automata -- the hodgepodge machine -- was modified to best match the chemical trajectory observed in the Belousov-Zhabotinsky reaction. Noise introduces watersheding of the central regular pattern into the circular target pattern. This article analyzes influences of the neighborhood and internal excitation kinds of noise. We have found that configurations of ignition points, which give circular waves -- target patterns, occur only in the interval of the neighborhood excitation noise from 30\% to 34\% and at the internal excitation noise of 12\%. Noisy hodgepodge machine with these parameters is the best approximation to the experimental Belousov-Zhabotinsky reaction. 
\keywords{cellular automaton, hodgepodge machine, noise}
\end{abstract}

\section{Introduction}
\label{intro}

Properties of multilevel cellular automata~\cite{Vanag,Gerhardt} are only sparsely examined in the contemporary literature. Hiltemann has shown that the state trajectory critically depends on the number of available levels \cite{Hiltemann2008} and we have concluded that there are a-few-level automata and true multilevel automata \cite{Stysetal2015}. The latter have a sufficient number of levels to allow the system's behavior to depend only on the internal evolution rule and not on the number of levels. The border between a-few and true multilevel automata was examined only for the square Moore neighborhood and was found to be 24~\cite{Hiltemann2008}.

The hodgepodge machine is a type of multilevel cellular automaton which mimics the final phase of the Belousov-Zhabotinsky (B-Z) reaction well. We have modified the model in order to start from a few ignition points as it is observed in the experiment \cite{Stysetal2015,Stysetal2016}. This enabled us to examine the properties of ignition points as well and the early phases of the trajectory.

A course of the hodgepodge machine simulation which is qualitatively similar to the observation -- i.e., contains a phase of large center structures, octagons filled by a complicated cross-like structure, eventually overcome by the mixture of spirals and waves -- was observed \cite{Stysetal2015} which invoked interpretation in the terminology of discrete dynamic networks. We have shown \cite{Stysetal2016} that the condition for existence of this qualitative dynamic trajectory is the asymmetry of the ignition unit when two neighboring points are needed for evolution of the wave. When the noise is introduced, the central octagonal structure is smeared into near-circular structures off and spirals were formed in the final stage of the simulation.

A trajectory of the noisy hodgepodge machine starts with an early phase of dense waves. Further, an intermediate phase, during which a symmetric structure in the center of the evolving waves is formed, follows \cite{Stysetal2015}. When two central structures coincide in proper geometry, spirals are formed. These spirals are eventually surrounded by a system of less dense waves which generate new spirals. Finally, the whole simulation canvas is covered by a mixture of spirals and waves, which closely imitates the observation of the final structure in the B-Z experiment. A clear analogy to the stages observed in the experiment is obvious.

The introduction of noise leads to the collapse of the central structure and formation of near circular concentric waves. As well, though the noise-free process would not lead to the formation of spirals and wave fragments -- apparently very stable structures frequently observed in the Nature, the noise induce their formation \cite{Stysetal2016}.

The near circular structures are similar to the target patterns observed in the beginning of the experiment \cite{Stysetal2016}. The later phase -- the mixture of spirals and waves -- is the ergodic structure observed in the basin of attraction and is similar to the final stage of the experiment. The main differences in the experiment are a lower regularity of spirals and waves and a slow decay of their borders due to the depletion of the reactants. The model which is described in this article is called a noisy hodgepodge machine (NHM). 

Based on the hodgepodge machine simulation, we proposed \cite{Stysetal2016} that the reaction mixture in the Petri dish is divided into a 2D grid of cells which is a consequence of nonlinear processes -- a  chemical analogy of the thermal process forming B\'{e}nard cells \cite{Benard}. The structure of the grid is unknown which limits our ability to properly describe, i.e. to model, the process. Also, precise parameters and identity of all individual chemical reactions are under constant dispute. The introduction of noise is an attempt to overcome this lack of knowledge.

In this article, we examine other details of the influence of noise on the outputs from the NHM modelling the B-Z reaction and discuss its relevance to the experiment. We found that there are certain combinations of parameters at which the transition from circular patterns to spirals and waves never occurs.

\section{Methods}

The B-Z reaction was performed as described by Dr. Jack Cohen \cite{Cohen_kit}. The differences were in the diameter of the Petri dish -- 200 mm instead of 90 mm -- and in very gentle mixing (1400 rpm). The experiment on a re-started B-Z reaction was performed by a manual shaking of the reaction vessel after reaching the state of dense waves. The photos of course of the experiment were taken in the time interval of 2 s and consists of 9 cycles of lengths of 48, 25, 44, 24, 25, 18, 11, 15, and 32 images, respectively.

The NHM model of the B-Z reaction is the same as described in \cite{Stysetal2016}:

In order to numerically implement a hotchpotch model, we have adopted an approach based on the version of Wilensky NetLogo model \cite{Wilensky2003}. In our case, the model was simulated on a square 1000 $\times$ 1000 grid. After a random setup of the space distribution of initial centers of $state(t = 0) \in [0, maxstate]$ as
\begin{equation}
state(t=0) = \mbox{random-exponential}[(maxstate+1) \times meanPosition]\,
\label{Eq1}
\end{equation}
where $maxstate$ = 200 is the maximally achievable number of levels of the cell state. Multiplication of each cell state by the $meanPosition$ of the exponential
distribution ensured that the simulation started with a small number of the ignition points. The model at each time step $t+1$ may proceed in four possible ways:
\begin{enumerate}
\item When a cell is at the $state(t) = 0$, so-called {\em quiescent}, it may be ``infected'' by surrounding cells according to the equation
\begin{equation}
state(t+1) = (1+PTN)\times\left[\mbox{precision}\left(\frac{a}{k_{1}}\right) +\mbox{precision}\left(\frac{b}{k_{2}}\right)\right],
\label{Eq2}
\end{equation}
where $a$ and $b$ is a number of cells at the $state \in (0, maxstate)$ and $state = maxstate$, respectively, $k_{1}$ = 3 and $k_{2}$ = 3 are characteristic constants of the process.
\item When a cell is at the $state(t) \in (0, maxstate)$, its new state is calculated as
\begin{equation}
state(t+1) = \mbox{precision}\left[(1+IEN)\times \frac{state(t) + \sum_{n=1}^{8} state_{n} (t)}{a + b +1} + (1+EEN)\times g\right],
\label{Eq3}
\end{equation}
where $state_{n} (t)$ is a state of the $n$-th cell in the Moore neighbourhood, which directly surrounds the examined cell, and $g$ = 28 is another arbitrary constant.
\item When a cell is at the $state(t) > maxstate$, then
\begin{equation}
state(t+1) = maxstate.
\label{Eq4}
\end{equation}
\item When a cell achieves the $state(t) = maxstate$, then
\begin{equation}
state(t+1) = 0.
\label{Eq5}
\end{equation}
\end{enumerate}

In Eq. \ref{Eq2}-\ref{Eq3}, precision of 10 decimal points were allowed for setting of the state and flat noise up to defined level was added to each of the processes. The individual noises are named
\begin{itemize}
\item the phase transition noise ($PTN$) which affects the transition from the state 0 to the first non-zero state,
\item the internal excitation noise ($IEN$) which affects the change of the state due to processes occurring inside the cell, i.e. the $g$ constant, and
\item the neighborhood (external) excitation noise ($EEN$) which affects processes related to the values of neighboring cells
\end{itemize}
and their influences on the NHM were tested by the systematic change of their values. Examples of qualitatively different cases are described in detail.

\section{Results and Discussion}

\subsection{Formation of target patterns}

\begin{figure}
\centering
{\includegraphics[width=\textwidth] {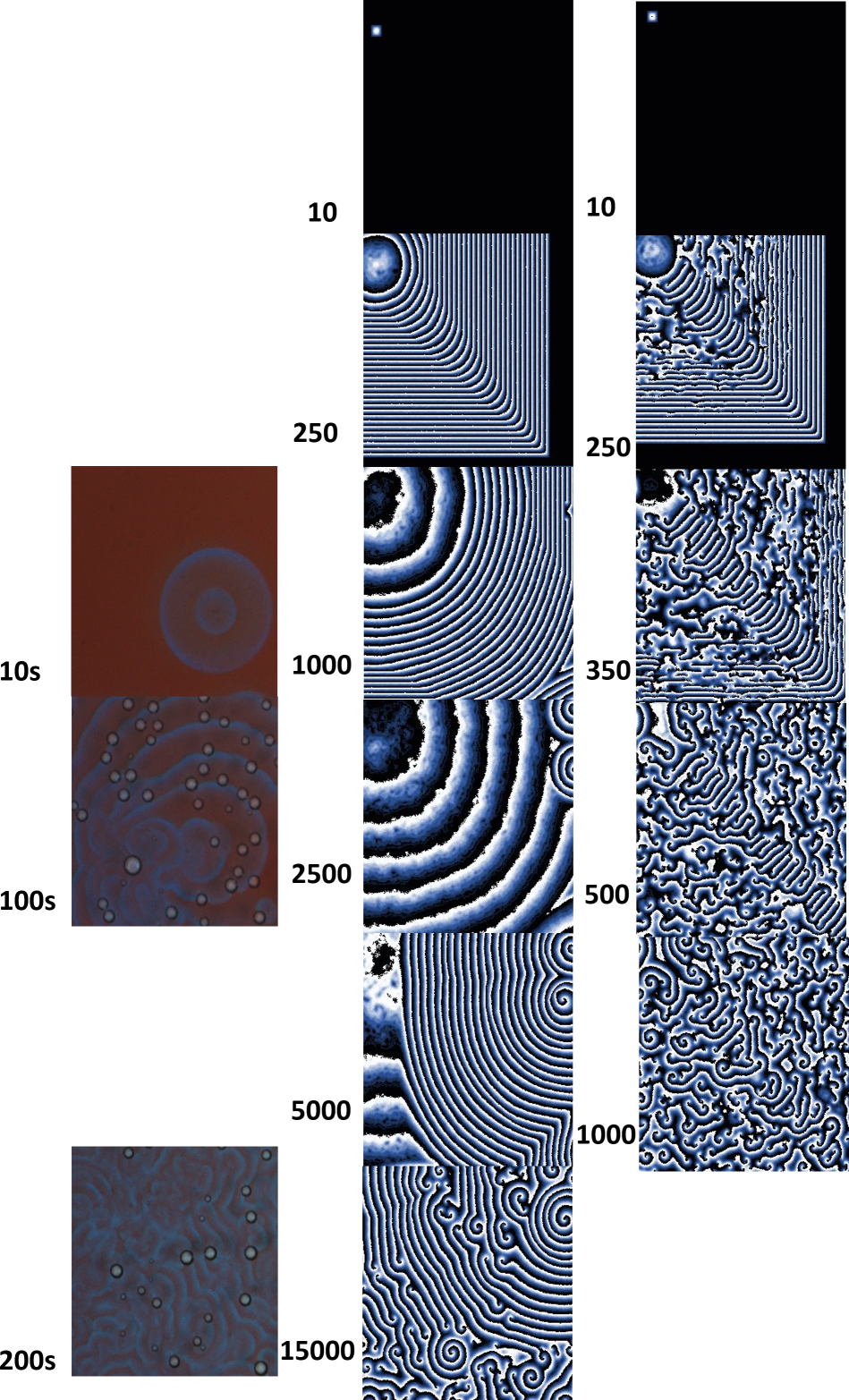}}
\caption{Comparison of a Belousov-Zhabotinsky experiment (\textit{left column}) with hodgepodge machine simulations at the external and internal excitation noise of 30\%--12\% (\textit{middle column}) and 30\%--16\% (\textit{right column}), respectively. Red-coded numbers correspond to the steps of the experiment and simulation, respectively.}
\label{fig:Experiment_two_simulations}
\end{figure}

Typical courses of the experiment and simulations are sketched in Fig. \ref{fig:Experiment_two_simulations}. In the experiment (\textit{left column}), the initial phase of square waves is not detected -- possibly, it is not observable in the spectral region which is detected by a color camera. As proposed in Section \ref{intro}, in this initial (lag) phase, partly regular grid on which the process occurs is formed. This phase takes about minutes and its duration depends on the method and time of mixing. After that, circular target patterns evolve from a few ignition points (\textit{10 s} in Fig. \ref{fig:Experiment_two_simulations}). The number of ignition points and frequency of circular waves depends sensitively on experimental conditions. Each degree of temperature changes the course of the reaction significantly. After the period of free evolution of circular waves, evolution of denser, less regular waves starts (\textit{100 s}). The dense waves typically arise at the border of the vessel, but, as in the case depicted in Fig. \ref{fig:Experiment_two_simulations}, they increasingly occur also in places of irregularities, mainly microbubbles. The structure observed in such centers resembles that of the first layer around the double spiral observed in the simulation. In the final -- ergodic -- state (\textit{200 s}), there is a mixture of waves and spirals which very slowly become homogenous as the chemical energy decays.

The simulation was designed to resemble the experiment as closely as possible. While the phase transition noise did not influence the results, the neighborhood excitation noise and the internal excitation noise decide together on properties of observed patterns. The initial state of the simulation is dominated by near-square patterns emanating from a single center (\textit{steps 10 and 250} in the \textit{middle} and \textit{right column}). Despite the fact that the central structures eventually circularized, the waves remained square until the central structure was broad enough to accommodate two circular waves (e.g., \textit{step 2500} in the \textit{middle column}). We have never been able to simulate more than 8 types of target pattern waves. Their frequency decreased with time -- waves became broader. When two dense waves merge, double spirals evolve and form a center for a new generation of dense waves -- the distance of the waves is 12 instead of 7 pixels. This eventually lead to the formation of the mixture of spirals and waves similar to that observed in the experiment.

The simulation with the external and internal excitation noise of 30\% and 12\%, respectively (\textit{middle column}) is that one which approximates the experiment best. In this case, we observe only one double spiral formation. When one of the noises is too high or \textbf{too low}, e.g., at external and internal excitation noise of 30\% and 16\%, respectively,  the number of double spirals is very high already in the phase of evolution of square waves (\textit{step 250} in \textit{right column}) and the circular waves -- target patterns -- will be never formed.

Among the most obvious differences between the experiment and this simulation belong that,
\begin{itemize}
\item in the experiment, the circularity of target patterns is nearly ideal, while, in the simulation, the waves are undulated and
\item in some geometrical arrangements of initial ignition centers, we do not observe formation of spirals and final dense waves. 
\end{itemize}

\subsection{Re-shaking experiment}

\begin{sidewaysfigure}
\centering
{\includegraphics[width=\textwidth] {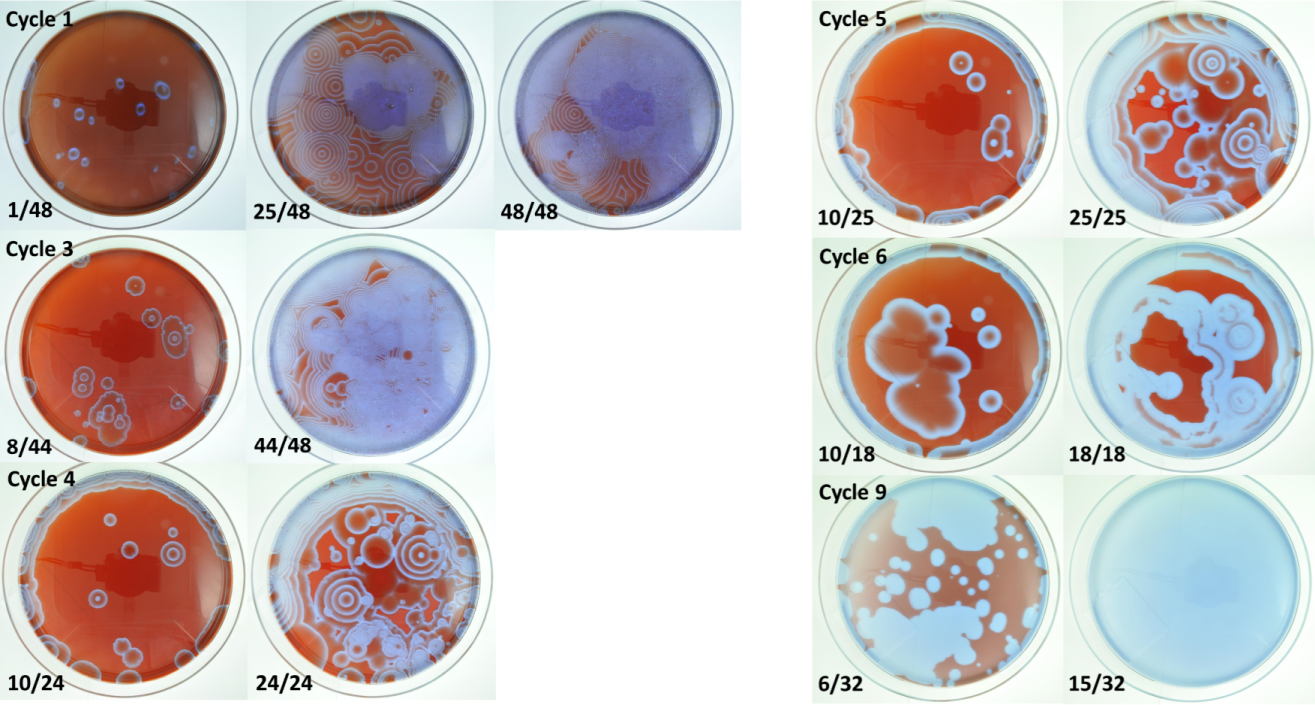}}
\caption{Re-start of the B-Z reaction (9 cycles). The number ratio X/Y means the X-th image from a Y-image series.}
\label{fig:Re-shaking_first_phase}
\end{sidewaysfigure}

Fig. \ref{fig:Re-shaking_first_phase} shows the course of the experiment on re-started B-Z reaction. The process (\textit{cycle 1} in Fig. \ref{fig:Re-shaking_first_phase}) started by the evolution of circular waves. Each sub-experiment was stopped after reaching a phase of dense waves and the reaction vessel was re-shaken. This process was repeated 9 times. Upon re-shaking, the waves gradually lost regularity and became thicker,  the diameters of target patterns increased (\textit{cycle 3} in Fig. \ref{fig:Re-shaking_first_phase}) and the evolution of the waves at the vessel's border (\textit{cycle 4} in Fig. \ref{fig:Re-shaking_first_phase}). Similar phenomena were observed in a Petri dish of a smaller diameter and we attribute them to the change of spatial relations and scales. Further thickening of waves (\textit{cycle 5} in Fig. \ref{fig:Re-shaking_first_phase}) led eventually to merging of circular waves (\textit{cycle 6} in Fig. \ref{fig:Re-shaking_first_phase}) up to a complete filling of circular waves' centres (\textit{cycle 9} in Fig. \ref{fig:Re-shaking_first_phase}). The next mixing did not lead to re-formation of red-coloured state.

%\begin{figure}
%\centering
%{\includegraphics[width=0.66\textwidth] {Reshaking_2nd_phase.png}}
%\caption{Follow-up Fig. \ref{fig:Re-shaking_first_phase}.  The number ratio X/Y means the X-th image from a Y-image series.}
%\label{fig:Re-shaking_second_phase}
%\end{figure}

In the early phase and, namely, upon gentle mixing (as shown in Fig. \ref{fig:Experiment_two_simulations}) the circular waves are highly regular. At later stages, upon re-shaking, the wavefronts become undulated and more similar to those observed in the simulation using NHM. Finally, the waves thickened to the extent that the formation of structures was no more possible.

This experiment demonstrates that the depletion of reactants does not change the shape of observed waves and their course (order) but causes thickening of the travelling waves and shortens time to reaching the ergodic state.

\subsection{Levels of noise determine the structure}

The course of the simulation depends on the level of noise. As noted in Section \ref{intro}, the formation of circular structures in the NHM is based on noise-induced breakage of the regular symmetrical structure which is formed in the center emanating waves in the noise-free hodgepodge machine \cite{Stysetal2015}.

\begin{figure}
\centering
{\includegraphics[width=\textwidth]{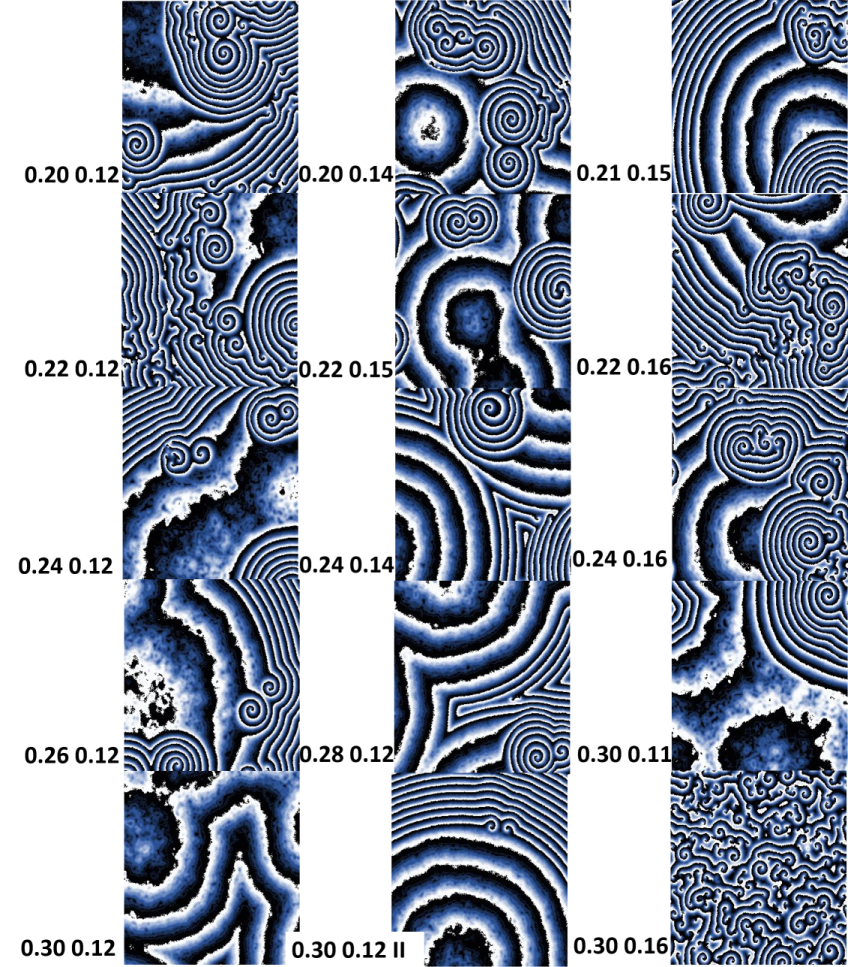}}
\caption{Sections of steps 1600 of the simulations at different levels of external (the first number) and internal (the second number) excitation noise. The Roman numeral II  (\textit{bottom, middle}) denotes the second experiment. Only at the internal and external (neighborhood) excitation noise of 30\% and 12\% (\textit{bottom, left}), respectively, mutual geometries of initial ignition points for which no spirals were formed were found. At higher density of ignition points and different geometries, spirals were formed even in these combinations of noises.}
\label{fig:step1600_simulation_increase_of_noise}
\end{figure}

In Fig. \ref{fig:step1600_simulation_increase_of_noise} we show a sketch of the research on the increase of neighborhood and internal kinds of excitation noise. Most cases gave a typical trajectory as shown in Fig. \ref{fig:Experiment_two_simulations}. Images in Fig. \ref{fig:step1600_simulation_increase_of_noise} show sections of the 1600$^{th}$ step of the simulation, where the spiral-based structures prevail over the central circular target pattern. We observed some remnants of the circular structures followed by spirals and waves evolving around them. However, both central circular structures and systems of spirals and waves slightly differ. The exception occured at neighborhood and internal excitation noise of 30\% and 12\%, respectively (\textit{bottom left} and \textit{bottom middle}), where, in some cases, we did not observe any spirals. In contrast, the combination of neighborhood and internal excitation noise of 30\% and 16\% (\textit{bottom right}) resulted in the fast evolution of spirals and waves which prevented the formation of circular waves.

\begin{figure}
\centering
{\includegraphics[width=.8\textwidth] {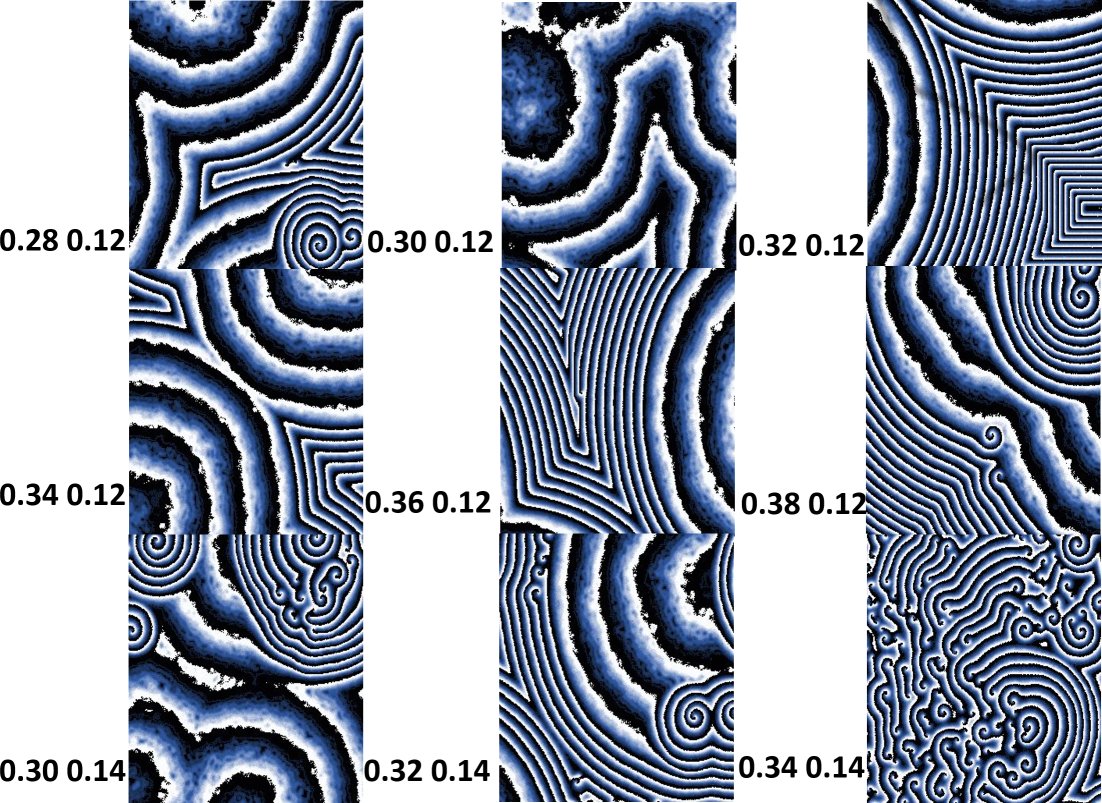}}
\caption{Steps 1600 of the simulations at different levels of external (the first number) and internal (the second number) excitation noise.}
\label{fig:step1600_simulation_0_14_0_12}
\end{figure}

In the next simulation series (Fig. \ref{fig:step1600_simulation_0_14_0_12}), the range of noise in which no spirals are formed was in detail examined and was found to be, at some geometric configurations of the ignition points, 12\% for the internal excitation noise and in the range from 30\% to 34\% for the neighborhood excitation noise (from \textit{upper middle} to \textit{middle left}). At the same internal excitation noise, a lower neighbourhood excitation noise (e.g., 28\% in \textit{upper left}) was too low to stabilize the circular structure of the spirals, while a higher level (e.g., 36\% in \textit{middle middle} and 38\% in \textit{middle right}) was too high to do so. At these conditions we observed the formation of spirals. The outputs of the simulations differed only in the length of the trajectories until the simulation canvas was fully filled by circular structures. The lengths of the segments of individual trajectories depended on both the levels of noise and the mutual positions of ignition points. As stated by Hiltemann \cite{Hiltemann2008}, the complete examination of all starting configurations require approx. 10$^{20}$ experiments and it is thus impossible to test all of them. At internal excitation noise higher than 12\% (\textit{bottom row} in Fig. \ref{fig:step1600_simulation_0_14_0_12} and respective images in Fig. \ref{fig:step1600_simulation_increase_of_noise}), it was not possible to find levels of neighborhood excitation noise at which only circular structures evolve.

\section{Conclusions}

In the B-Z experiments, the target circular waves are always overcome by dense waves and spirals. Dense waves are typically evolving at the border of a Petri dish due to the non-idealities of the spatial geometry. Spirals which evolve from the center origin from a microbubble -- again, from a spatial inhomogeneity.

The re-shaked experiment excludes any simple chemical interpretation of the decay of observed structures. It is not the depletion of chemicals which leads to the transformation of circular waves -- target patterns -- to dense waves and, finally, to the mixture of spirals and waves. In the wide range of concentrations, when the thickness of waves is not broader than the diameter of the Petri dish, the general course of the B-Z reaction is qualitatively identical. \textit{The self-organization in the B-Z reaction is a process which is separated from a concrete chemical reaction.} It entitles to search for a model of self-organization which would describe the reaction and ignore the actual chemical process.

In the simulation presented in this article, it has been found that, at certain configurations of ignition points, there is a lower and upper limit of the noise at which the whole simulation canvas is filled by circular structures -- target patterns and the phase of spirals and waves never occurs. It is at the combination of neighborhood (external) excitation noise from 30\% to 34\% with the internal excitation noise of 12\%. The spatial inhomogeneity which lead to the evolution of spirals and waves at unfavourable conditions is not properly described by the model. However, at certain combinations of the geometry of ignition points, spirals are formed even in this interval.

Differences in structures and dynamics shown in the re-shaking experiment (Fig. \ref{fig:Re-shaking_first_phase}, the undulation of circular waves, thickening, doubling of wavefront, etc., indicate that there are numerous individual processes which play a role in the formation of the pattern of B-Z reaction. All these processes have rates comparable to the bottleneck process which determines time of the reaction. There is no known experimental procedure for identification of these processes -- we know a lot of chemicals but we do not know which breakage of individual chemical bond or diffusion constant is the bottleneck one. 

We conclude that the noisy hodgepodge machine -- NHM -- is the simplest approximation which provides all stages observed in the experiment and indicates basic geometrical and kinetic rules. The ratio of two slowest processes close to 7:2 and the need of the asymmetry of ignition unit seem to be a well confirmed. However, in reality, we observe a process which combines a number of individual processes of similar rate and spatial extent which form a robust process which dominates over all competing processes.

\subsubsection*{Acknowledgments.} This work was financially supported by CENAKVA (No. CZ.1.05/2.1.00/01.0024), CENAKVA II (No.LO1205 under the NPU I program) and The CENAKVA Centre Development (No. CZ.1.05/2.1.00/19.0380).


\begin{thebibliography}{4}
\bibitem{Vanag} Vanag, V.K.: Study of Spatially Extended Dynamical Systems Using Probabilistic Cellular Automata. Physics--Uspekhi 42(5), 413--434 (1999)
\bibitem{Gerhardt} Gerhardt, M., Schuster, H., Tyson, J.J.: A Cellular Automation Model of Excitable Media Including Curvature and Dispersion. Science 30, 1563--1566 (1990)
\bibitem{Hiltemann2008} Hiltemann, S.: Multi-coloured cellular automata. Bachelor thesis, Erasmus Universiteit, Rotterdam, Netherlands (2008)
\bibitem{Stysetal2015} \v{S}tys, D., N\'{a}hl\'{i}k, T., Zhyrova, A., Rycht\'{a}rikov\'{a}, R., Pap\'{a}\v{c}ek, \v{S}., C\'{i}sa\v{r}, P.: Model of the Belousov-Zhabotinsky Reaction, in press in LNCS 9611, Kozubek, T., Blaheta, R., \v{S}\'{i}stek, J., Rozlo\v{z}n\'{i}k, M., \v{C}erm\'{a}k, M. (Eds.), available at http://arxiv.org/pdf/1507.08783v2.pdf
\bibitem{Stysetal2016} \v{S}tys, D., Jizba, P., Zhyrova, A., Rycht\'{a}rikov\'{a}, R., \v{S}tys, K. M., N\'{a}hl\'{i}k, T.: Multi-State Stochastic Hotchpotch Model Gives Rise to the Mesoscopic Behaviour in the Non-Stirred Belousov-Zhabotinky Reaction, submitted to Chaos, available at http://arxiv.org/pdf/1602.03055v1.pdf
\bibitem{Benard} Getling, A.V.: Rayleigh-B\'{e}nard Convection: Structures and Dynamics. World Scientific (1998)
\bibitem{Cohen_kit} Cohen, J., \url{http://drjackcohen.com/BZ01.html}
\bibitem{Wilensky2003} Wilensky, U. (2003). NetLogo B-Z Reaction model. http://ccl.northwestern.edu/netlogo/models/B-ZReaction. Center for Connected Learning and Computer-Based Modeling, Northwestern University, Evanston, IL.
%\bibitem{Krapivsky} Krapivsky, P.L., Redner, S., Ben-Naim, E.: A Kinetic View of Statistical Physics. Cambridge University Press, Cambridge (2010)
\end{thebibliography}
\end{document}